\documentclass[preprint,11pt]{elsarticle}
\biboptions{sort,sort&compress,square,numbers}

\usepackage{amsmath}
\usepackage{graphics}
\usepackage{graphicx}

\usepackage{anysize}
\papersize{29.7cm}{21.0cm} 
\marginsize{2cm}{2.5cm}{2.5cm}{2.5cm}
\begin{document}



\pagestyle{myheadings} 

\begin{abstract}
Tourism is a complex dynamic system including multiple actors which are related each other composing an evolving social network. 
This paper presents a growing bipartite network model that explains the rise of the supply network in a tourism destination from the beginning phases of development. The nodes are the lodgings and services in a destination and a link between them appears if a representative tourist hosted in the lodging visits/consumes the service during his/her stay. The specific link between both categories are determined by a random and preferential attachment rule. The analytic results show that the long-term degree distribution of services follows a shifted power-law distribution. The numerical simulations show slight disagreements with the theoretical results in the case of the one-mode degree distribution of services, due to the low order of convergence to zero of X-motifs. The model predictions are compared with real data coming from a popular tourist destination in Gran Canaria, Spain, showing a good agreement between numerical and empirical data. 

\end{abstract}

\begin{frontmatter}

\title{\textbf{An evolving model for the supply network in a tourism destination}}

\author[rvt]{Juan M.~Hern{\'a}ndez\corref{cor1}
} \ead{juan.hernandez@ulpgc.es}

\author[focal]{Christian~Gonz{\'a}lez
}
\ead{christian.gonzalez@ulpgc.es}

\cortext[cor1]{Corresponding author}
\address[rvt]{Department of Quantitative Methods in Economics, Institute of Tourism and Sustainable Development (TIDES), University of Las Palmas de Gran Canaria, c/Saulo Tor{\'o}n s/n, 35017, Las Palmas, Spain}
\address[focal]{Department of Quantitative Methods in Economics, University of Las Palmas de Gran Canaria, c/Saulo Tor{\'o}n s/n, 35017, Las Palmas, Spain}

\end{frontmatter}

\section{Introduction}

In the last two decades, the complex networks perspective has been widely used to analyze diverse socioeconomic phenomena, 
such as professional collaborations  \cite{Newman2001d,Goyal2006a,Watts1998}, international trade networks \cite{Serrano2003,Garlaschelli2005,Fagiolo2009}, business networks \cite{Souma2003,Inaoka2004,Mizuno2014} and also tourism \cite{Scott2008,Romeiro2010,Shih2006,Baggio2007,
Baggio2010,Miguens2008,Grama2014,Sainaghi2014,Smallwood2012,Baggio2016}. 
The case of tourism is noteworthy. This is a geographical, economic and social phenomenon which consists on the temporary movement of some people (tourists) from origin countries to certain destinations, where they spend time enjoying some services (visiting natural or man-made monuments, restaurants, going shopping and other activities). The big amount of agents involved in this industry (visitors, lodgings, air carriers, local suppliers, services in destination, etc.) and the strong interdependence among them make this industry to be considered a complex adaptive system and susceptible to be analyzed using complex network methodology \cite{Baggio2014}.\par

Nevertheless, the application of the complex networks approach on tourism research is still limited as compared to other fields. One of the reasons is the sample size, which is usually small in the real networks analyzed so far. This is due to the common methodology applied for the data collection, based on interviews of agents involved, which makes that obtaining samples larger than $10^2$ nodes is monetary and time costly 
\cite{Scott2008,Romeiro2010,Shih2006,DaFontouraCosta2009,
Baggio2007,Grama2014,Sainaghi2014,Smallwood2012}. Regarding the issues explored, many of the previous contributions make statistical analyses to characterize the topological structure of a specific tourism network  \cite{Baggio2007,Miguens2008,Grama2014} or modularity analysis to detect communities \cite{Baggio2016,DaFontouraCosta2009,Scott2008}. Other papers go further and apply the superedges approach to tourism data \cite{DaFontouraCosta2009}, study the knowledge diffusion in the supply network \cite{Baggio2010}, the influence of network topology on social capital of hotels \cite{Sainaghi2014} or the movement pattern of tourists in destinations \cite{Smallwood2012}.\par 

Looking at the specific definition of the tourism network used, some of the contributions above study the supply-side network, where nodes contains stakeholders, such as hoteliers and travel agencies, which are linked through business associations or website links \cite{Baggio2010,Grama2014,Scott2008}. Other networks include locations who are linked by travelers \cite{Miguens2008,Shih2006}. Unlike to the rest, Smallwood et al. \cite{Smallwood2012} analyze the tourist network combining the demand (tourist) and supply-side (activities/attractions).\par

To the extent of our knowledge, the application of complex networks to analyze the growth dynamic of the supply network in a tourism destination is still unexplored. However, this is a major issue in tourism research, since it would allow identifying the main forces leading to a destination from the initiating to mature stage and provide key factors about its next future evolution.\par

In this regard, the adaptation of evolving models of bipartite networks to the tourism context is an useful way to represent the phenomenon. Evolving models has been theorized for one-mode networks \cite{Barabasi1999,Dorogovtsev2013} and extensions to several cases of bipartite networks have been also proposed \cite{Ramasco2004,Noh2004,Nacher2009,Tian2012,Zhang2013, Zhang2015b,Qiao2016}. Following the same basis of one-mode evolving networks, these models assume that new nodes of both categories appear in every time step. The major difference among them is the specific combination of random and preferential attachment (PA) rules to link new and old nodes. For example, Ramasco et al. \cite{Ramasco2004} assume new nodes are attached following an unidirectional combination of these rules (e.g., new movies select an amount of old and new actors), while Tian et al. \cite{Tian2012} and Zhang et al. \cite{Zhang2013} assume bidirectional PA rule for selecting links between nodes from two categories, allowing rewiring which follows PA and/or random attachment rules. The latter models have been extended by Zhang et al. \cite{Zhang2015b} including weight in the links. Alternatively, Noh et al. \cite{Noh2004} create a growing model to represent the community growth through group membership, assuming that the join of a new member to an old group or creation of new groups is produced according old member's degree and random rules. Finally, Nacher et al. \cite{Nacher2009} adapt a bipartite growing network to represent protein-domain network including a copy mechanism and random rule for new nodes. As it was the case for one-mode distributions, the long-term degree distribution in the contributions above varies from shifted power-law to exponential distribution, depending on the specific weight of PA or random attachment rules.\par

This paper proposes an evolving model of bipartite network to represent the development of the supply network in a tourism destination. The model explains how the supply components in the destination grow and are connected each other through tourists' behavior. Specifically, the model describes a social network where lodgings are linked with services. A link appears if a representative tourist staying in a certain lodging enjoys a service. The propensity to enjoy a specific service follows a combination of preferential and random attachment rules.  
In the long term, services degree follows a shifted power-law or exponential distribution. The model estimations are tested with real data coming form a tourist area in Gran Canaria, Spain, collected from recommendations made by visitors in the booking web-site tripadvisor.com. The results show good agreement between theoretical, numerical and real data for the service degree distribution. The one-mode projection does not show so good fit results between numerical and theoretical data, due in part to the low order of convergence of the degree distribution to the asymptotic state. Model limitations and future research are commented in the conclusions section. 

\section{The model}

An evolving network is proposed to explain the supply growth in a certain destination. We assume two categories of nodes, lodgings ($H$) and services ($S$). Lodgings include the accommodation units in the destination, such as hotels, apartments, Bed\&Breakfast, etc., and services include all the activities (visiting monuments, restaurants, etc.) that the tourist can make during the stay. The model shows how the links between elements of these categories appear along time until achieving an asymptotic network structure.\par

We start at time $t_0$ with an initial bipartite network, which is represented by a triple $G=(H_0,S_0,L_0)$, where $H_0$ and $S_0$ represents the initial lodgings and services in the destination, with $card(H_0)=H^0$ and $card(S_0)=S^0$. We assume that $S^0=mH^0$, with $m \geq 1$. The network only includes links between elements of $H_0$ and $S_0$. We assume the existence of a representative tourist in every lodging. A link between a lodging $i \in H_0$ and service $j \in S_0$ appears if the representative tourist of lodging $i$ visits service $j$ during his/her staying in the destination. Translated into real visits, a link would mean that a high enough number of tourists hosted in $i$ visits service $j$. So, the links are unweighted and undirected. Additionally, we assume them permanent along time, which means that, once a service is enjoyed by a high enough amount of tourists staying in a certain lodging, the preference for this service is maintained by successive guests. By simplicity, we also assume that every lodging includes exactly $c$ links to services, with $c \geq 1$.\par

The bipartite network grows as follows. At any time $t > t_0$, a new lodging and $m$ new services are created in the destination. We assume that the representative tourist of all new hotels visit $c$ different services, selected between the old and new ones, following the rule: a) A percentage $\phi \in [ 0,1 ] $ of them at random; b) The rest by linear PA according to service's degree, so the higher the current links to a service, the higher the probability to be visited by tourists in the new lodging. This tourist's behavior rule is sustained by previous theoretical and empirical studies on within destination movements of tourists \cite{Lew2006,McKercher2008}.
\par

We note $s_j(t)$ the degree of service $j$ at time $t$. Next, we derive some equations to describe the evolution of $s_j$. Previously, we note $S(t)=S^0 +tm$  and $H(t)=H^0 + t$ the number of services and lodgings at time $t$, respectively. Since $S^0=mH^0$, we have $S(t)=mH(t)$. The relationship between the average degree of lodgings and services is as follows: $<h>^t=c =\displaystyle{\frac{\sum_i h_i(t)}{H(t)} = m \frac{\sum_j s_j(t)}{S(t)}}=m <s>^t$, so $<s>^t=\displaystyle{\frac{c}{m}}$. Naming $L_j^t$ the event that service $j$ is linked by a new lodging at time $t+1$, the probability of this event is
\begin{equation}
P\left[ L_j^t \right]= c\frac{(1-\phi)s_j(t)}{\sum_js_j(t)} +c\frac{\phi}{S(t)}. \label{prob1}
\end{equation}
The first term on the right-hand side in (\ref{prob1}) 
is the probability that one of the $c$ links coming from the new lodging goes to service $j$ via PA, so it depends on its current degree $s_j(t)$. The second term is the probability that service $j$ be attached by one of these new links at random. Applying the relationship between $S(t)$, $H(t)$ and the average degrees $<s>^t$ and $<h>^t$ described above, the probability (\ref{prob1}) can be simplified into
\begin{equation}
(1-\phi)\frac{s_j(t)}{H^0+t} +\frac{c \phi}{m(H^0+t)} = \frac{(1-\phi)m s_j(t) + c \phi}{m(H^0+t)}. \label{prob2}
\end{equation}
Now, assuming $t>>H_0$ and using the continuous time approximation, the evolution of $s_j$ can be described by
\begin{equation}
\frac{\partial s_j}{\partial t}=\frac{(1-\phi)m s_j + c \phi}{m t}, \label{evsjt}
\end{equation}
subject to the initial condition $s_j(t_j)=0$, where $t_j$ is the time when service $j$ is included in the network. The solution of (\ref{evsjt}) is
\begin{equation}
s_j(t)=\frac{c \phi}{m (1-\phi)} \left(  \frac{t}{t_j} \right)^{1-\phi} - \frac{c \phi}{m (1-\phi)}. \label{sol_evsjt}
\end{equation}
In order to derive the probability distribution of $s_j$ in the asymptotic limit ($t \rightarrow \infty$), we make use that the age of a random selected service $j$ at time $t$ follows an homogeneous distribution $p(t_j)=\frac{1}{H^0+t}$, since the number of services included in the network at every time is the same. Therefore, making some calculations on (\ref{sol_evsjt}), the cumulative probability for $s_j(t)$ can be written
$$
P\left( s_j(t)<s \right) = P\left( t_j > t \left( \frac{c \phi}{c \phi + m (1-\phi)s} \right)^\frac{1}{1-\phi} \right)=
1- \frac{t}{H^0+t} \left( \frac{c \phi + m (1-\phi)s}{c \phi} \right)^{-\frac{1}{1-\phi}}, \forall j \in S(t).
$$
Now, the distribution probability can be approximated by
$$
p(s) = \frac{\partial P\left( s_j(t)<s \right)}{\partial s} = \frac{t}{H^0+t} m(c \phi)^{\frac{1}{1-\phi}} \left( c \phi + m (1-\phi)s \right)^{-\frac{1}{1-\phi}-1}.
$$
Assuming $t$ sufficiently large, we have
\begin{equation}
p(s) \simeq  m(c \phi)^{\frac{1}{1-\phi}} \left( c \phi + m (1-\phi)s \right)^{-\frac{1}{1-\phi}-1}. \label{dist_s}
\end{equation}
Depending on the research field, this is called a shifted power-law \cite{Zhang2013} or Pareto distribution \cite{Feller1971,Ramsay2006}. In the extreme cases of parameter $\phi$, the distribution (\ref{dist_s}) fluctuates from a power-law $p(s) \sim s^{-\gamma}$, with $\gamma=2$ ($\phi \rightarrow 0$) to an exponential distribution ($\phi \rightarrow 1$). The heavy tails presented in case of adoption of pure PA rule are also produced for extreme values of parameters in other evolving bipartite network models \cite{Ramasco2004,Tian2012,Zhang2013}. An identical shifted power-law exponent than the BA model ($\gamma=3$) is obtained for $\phi = \displaystyle{\frac{1}{2}}$. The first moment is, as expected, $<s> = \displaystyle{\frac{c}{m}}$ and the second moment is $<s^2>=\displaystyle{\frac{c^2\phi}{\left( \phi - \frac{1}{2} \right)m^2}}$ if $\displaystyle{ \phi > \frac{1}{2}}$ and diverges in other case. \par

\subsection{Degree of the one-mode projection on services}

According to this projection, service $j$ and $j^\prime$ are linked if they are both linked to the same lodging in the bipartite network. In other words, a link appears between both services if they are visited by the representative tourist staying in one lodging. Let us define $k_j$ the degree of service $j$ in the one-mode projection. The relation between $k_j$ and $s_j$ is the following:
\begin{equation}
k_j=(c-1) s_j - q_j, \forall j \in S. \label{as1}
\end{equation}
The first term on the right includes all links to other services from a lodging connected to $j$. It is possible that the connection to other service $j^\prime$ is duplicated if those services ($j$ and $j^\prime$) are visited by tourists staying in two different lodgings ($i$ and $i^\prime$). In this case, services $\{j,j^\prime\}$ and lodgings $\{i,i^\prime\}$ form an X-motif, as it is illustrated in Figure 1. The second term $q_j$ represents the number of X-motifs which include service $j$. The degree $k_j$ is obtained by subtracting the duplicates to all links from service $j$.\par

\begin{figure}[!ht]
\centering
    \includegraphics[width=0.15\textwidth, height=0.2\textheight]{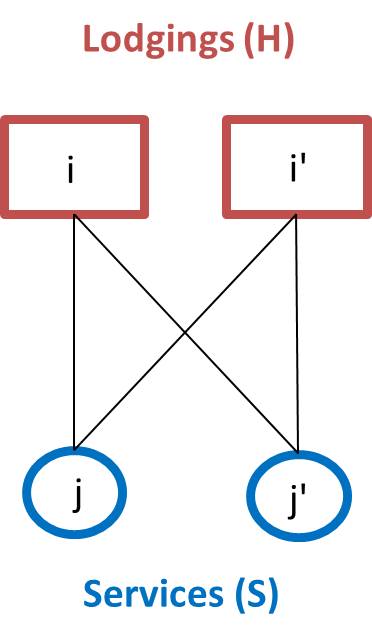}
    
\caption{X-motif between lodgings $\{i,i^\prime\}$ and services $\{j,j^\prime\}$.}
\end{figure}

It is expected than the number of X-motifs in the bipartite network tends to be a null-measure set as time evolves. In fact, this is proven in the next section. So, we can assume in (\ref{as1}) that $q_j(t)\simeq 0$ for large values of $t$ and the one-mode degree of service $j$ can be approximated by $k_j=(c-1) s_j, \forall j \in S$. Starting from the distribution (\ref{dist_s}), we can obtain the probability distribution $p(k)$ in the same way as in \cite{Ramasco2004}. Applying the change-of-variable formula, we have after some calculations
\begin{equation}
p(k)=\frac{1}{c-1}p\left( \frac{s}{c-1} \right)  \simeq  m(c (c-1) \phi)^{\frac{1}{1-\phi}} \left( c (c-1) \phi + m(1-\phi)k \right)^{-\frac{1}{1-\phi}-1}, \label{dist_k}
\end{equation}
which is again a shifted power-law distribution.\par

\subsection{Probability distribution of X-motifs}

Next, we show that the probability of finding an X-motif including service $j$ in the network at time $t$, $q_j(t)>0$, is null when $t \rightarrow \infty$. Additionally, we estimate its order of convergence to zero, which will be below 1. This low order of convergence makes that these motifs, neglected in the one-mode degree distribution estimation (\ref{dist_k}), influence on the fit of the theoretical model to numerical simulations for short time horizons.\par

We name $L_{jj^\prime}^t$ the event that service $j$ and $j^\prime$ links to a common lodging at any time before $t+1$. Below we present the probability that a lodging introduced at time $t+1$ links service $j$ and any other one such that both have been previously linked by other lodging, 
\begin{equation}
P\left[\cup_{j^\prime \neq j} (L_j^t \cap L_{j^\prime}^t \cap L_{jj^\prime}^t) \right] \simeq \sum_{j^\prime \neq j} P\left[ L_j^t \right] P\left[ L_{j^\prime}^t/L_j^t \right] P\left[ L_{jj^\prime}^t \right].\label{prob3a}
\end{equation}
For simplicity, we have disregarded the lower order terms of the probability on the left and assume that events $L_j^t \cap L_{j^\prime}^t$ and $L_{jj^\prime}^t$ are independent. The first two probabilities on the right hand side can be easily calculated from (\ref{prob1}), while the third one is
$$
P\left[L_{jj^\prime}^t \right] \simeq \sum_{r=max(t_j,t_{j^\prime})}^t P\left[ L_j^r \right] P\left[ L_{j^\prime}^r/L_j^r \right],\label{prob3b}
$$
where we have disregarded again the lower order terms. Therefore, making use of the continuous approximation, probability (\ref{prob3a}) can be calculated as follows, 
\begin{equation}
\begin{split}
\displaystyle{P\left[\cup_{j\neq j^\prime} (L_j^t \cap 
L_{j^\prime}^t \cap L_{jj^\prime}^t) \right]
\simeq }
\displaystyle{\int_0^t \left( \frac{c-1}{c} \right)^2 } & \displaystyle{\frac{(1-\phi)m s_j(t) + c \phi}{m(H^0+t)} \frac{(1-\phi)m s_j^\prime(t) + c \phi}{m(H^0+t)} }  \\ & \displaystyle{\times\left( \int_{max(t_j,t_j^\prime)}^t \frac{(1-\phi)m s_j(r) + c \phi}{m(H^0+t)} \frac{(1-\phi)m s_j^\prime(r) + c \phi}{m(H^0+t)} dr \right) dt_{j^\prime} }.
\label{prob3c}
\end{split}
\end{equation}
Now, we take into account only the higher-order terms and assume $t>>H^0$. Then, assuming $\phi \neq \frac{1}{2}$, (\ref{prob3c}) is approximated by\footnote{The results of the case $\phi = \frac{1}{2}$ are a continuous extension of results presented here.}

\begin{equation}
\begin{array}{l}
\displaystyle{ \frac{(c-1)^2(1-\phi)^4}{c^2} \frac{t^{-\phi}}{t_{j}^{1-\phi}} \left[ \int_{t_0}^{t_j} \frac{t^{-\phi}}{t_{j^\prime}^{1-\phi}} \left( \int_{t_j}^t \frac{r^{-2\phi}}{t_{j}^{1-\phi} t_{j^\prime}^{1-\phi}}  dr \right) dt_{j^\prime} + 
\int_{t_j}^{t}
 \frac{t^{-\phi}}{t_{j^\prime}^{1-\phi}} \left( \int_{t_{j^\prime}}^t \frac{r^{-2\phi}}{t_{j}^{1-\phi} t_{j^\prime}^{1-\phi}}  dr \right) dt_{j^\prime} \right]} \\ \displaystyle{ = \frac{(c-1)^2(1-\phi)^4}{c^2(1-2\phi)} \left( \frac{t_0^{-1+2\phi}}{1-2\phi} \frac{t^{1-4\phi}}{t_j^{2(1-\phi)}}- \frac{t^{-2\phi}}{t_j^{2(1-\phi)}} ln \left( \frac{t}{t_j} \right) -\frac{t_0^{-1+2\phi}}{1-2\phi} \frac{t^{-2\phi}}{t_j} \right) }.
\end{array}
\end{equation}
Again, using the continuous approximation, the growth of $q_j(t)$ can be described by the following differential equation, 
\begin{equation}
\frac{\partial q_j}{\partial t}=\frac{(c-1)^2(1-\phi)^4}{c^2(1-2\phi)} \left( \frac{t_0^{-1+2\phi}}{1-2\phi} \frac{t^{1-4\phi}}{t_j^{2(1-\phi)}}- \frac{t^{-2\phi}}{t_j^{2(1-\phi)}} ln \left( \frac{t}{t_j} \right) -\frac{t_0^{-1+2\phi}}{1-2\phi} \frac{t^{-2\phi}}{t_j} \right), \label{evqjt}
\end{equation}
where $q_j(t_j)=0$, since service $j$ cannot belong to any X-motif at the time it is inserted in the network. At this time $t_j$, only the new lodging can link service $j$. Equation (\ref{evqjt}) can be easily integrated. The solution is
\begin{equation}
q_j(t)=A \left[ a_1 \frac{t^{2(1-2\phi)}}{t_j^{2(1-\phi)}} + a_2 \frac{t^{1-2\phi}}{t_j^{2(1-\phi)}}ln \left( \frac{t}{t_j} \right) + a_3 \frac{t^{1-2\phi}}{t_j^{2(1-\phi)}} + a_4 \frac{t^{1-2\phi}}{t_j} + a_5 \frac{1}{t_j^{2\phi}} + a_6 \frac{1}{t_j} 
\right], 
\end{equation}
where $A=\frac{(c-1)^2(1-\phi)^4}{c^2} \geq 0$ and the rest of parameters are
$$
\begin{array}{lll}
\displaystyle{  a_1=\frac{t_0^{-1+2\phi}}{2(1-2\phi)^3}, } &\displaystyle{ a_2=-\frac{1}{(1-2\phi)^2}, }& \displaystyle{a_3=\frac{1}{(1-2\phi)^3}, }\\
\displaystyle{a_4=-\frac{t_0^{-1+2\phi}}{(1-2\phi)^3}, }& \displaystyle{a_5=\frac{t_0^{-1+2\phi}}{2(1-2\phi)^3},} &\displaystyle{ a_6=-\frac{2\phi}{(1-2\phi)^3}. }\\
\end{array}
$$
In order to obtain the probability distribution of $q_j$, we follow the same procedure than above. When $t>>t_0$, the cumulative probability is 
\begin{equation}
\begin{split}
P\left( q_j(t)<q \right)=P \left( \frac{q}{A} t_j^{2(1-\phi)} > a_1 t^{2(1-2\phi)}\right. & + a_2 t^{1-2\phi}ln \left( \frac{t}{t_j} \right) + a_3 t^{1-2\phi}  \\&\left.  + a_4 t^{1-2\phi}t_j^{1-2\phi} +  a_5 t_j^{2(1-2\phi)}  + a_6 t_j^{1-2\phi}  \right).
\end{split}
\label{probqj}
\end{equation}
A closed form function for this probability is unknown, but we can bound it by a known probability distribution, using that $t_0 \leq t_j \leq t$. We have to distinguish two cases: 

\begin{itemize}

\item Assume $\displaystyle{0<\phi < \frac{1}{2}}$. In this case, we have that 
$0 < 1-2\phi<2(1-2\phi)$ and $a_1, a_3, a_5$ are positive, the rest negative. Probability (\ref{probqj}) can now be bounded by 
\begin{multline*}
P \left(q_j(t)<q \right) \geq P \left( \frac{q}{A} t_j^{2(1-\phi)} > a_1 t^{2(1-2\phi)} \right. \left. + a_2 t^{1-2\phi}ln \left( \frac{t}{t_0} \right) + a_3 t^{1-2\phi} + a_4 t^{1-2\phi}t_0^{1-2\phi} \right. \\ \left.+  a_5 t^{2(1-2\phi)}  + a_6 t_0^{1-2\phi} \right) \end{multline*}
\begin{equation}
= P \left( t_j > \left(\frac{A}{q}\right)^\frac{1}{2(1-\phi)} \left( (a_1 +a_5) t^{2(1-2\phi)} + l.o.t \right)^\frac{1}{2(1-\phi)} \right) 
\simeq 1-\left(\frac{A(a_1+a_5)}{q}\right)^\frac{1}{2(1-\phi)} \frac{ t^{\frac{1-2\phi}{1-\phi}}}{H_0+t}.
\label{probqj3}
\end{equation}
Then, making the derivative of latter function in (\ref{probqj3}), we get that the probability distribution $p(q)$ first-order dominates the following probability distribution $\bar{p}(q)$, assuming $t>>H_0$, 
\begin{equation}
\bar{p}(q) = \frac{(Aa_1+ Aa_5)^\frac{1}{2(1-\phi)}}{t^{\frac{\phi}{1-\phi}}} q^{-1-\frac{1}{2(1-\phi)}}. 
\label{pqbar1}
\end{equation}
This distribution probability follows a power-law with exponent $-1-\frac{1}{2(1-\phi)}$. However, in the asymptotic limit ($t \rightarrow \infty$), the right-hand side of equation (\ref{probqj3}) converges to 1, so $P(q_j(t)<q) \simeq 1$, $\forall q>0$, with an order of convergence $0<\frac{\phi}{1-\phi}<1$. Therefore, the expected number of X-motifs is zero for $t$ sufficiently large.

\item Assume $\displaystyle{\frac{1}{2}<\phi<1}$. In this case, the exponents $2(1-2\phi)<1-2\phi<0$ and $a_4, a_6$ are positive, the rest negative. Then, probability (\ref{probqj}) can be bounded by 
\begin{multline*}
P\left(q_j(t)<q \right) \geq P \left( \frac{q}{A} t_j^{2(1-\phi)} > a_1 t^{2(1-2\phi)} + a_2 t^{1-2\phi}ln \left( \frac{t}{t_0} \right) + a_3 t^{1-2\phi} + a_4 t^{1-2\phi}t_0^{1-2\phi} \right. \\ \left.+  a_5 t^{2(1-2\phi)}  + a_6 t_0^{1-2\phi} \right)
\end{multline*}
\begin{equation}
= P \left( t_j > \left(\frac{A}{q}\right)^\frac{1}{2(1-\phi)} \left( a_6 t_0^{1-2\phi} + l.o.t \right)^\frac{1}{2(1-\phi)} \right)
\simeq 1-\left(\frac{A}{q}\right)^\frac{1}{2(1-\phi)} \frac{(a_6 t_0^{1-2\phi})^\frac{1}{2(1-\phi)}}{H_0+t}. 
\label{probqj2}
\end{equation}
Again, the probability distribution $p(q)$ first-order dominates the probability distribution $\bar{p}(q)$, which can be approximated by 
\begin{equation}
\bar{p}(q) = \frac{(Aa_6 t_0^{1-2\phi})^\frac{1}{2(1-\phi)}}{H_0+t}q^{-1-\frac{1}{2(1-\phi)}}. \label{pqbar2}
\end{equation}
We have the same power-law distribution with exponent $-1-\frac{1}{2(1-\phi)}$, but it degenerates for large $t$ to the value $q=0$ with an order of convergence $1$. So, in the long term the expected number of X-motifs (Figure 1) is a null-measure set and can be disregarded. 

\end{itemize}

Table 1 shows the analytic and numerical calculations of the order of convergence for the distribution $\bar{p}(q)$ and $p(q)$, respectively, given a range of values of the percentage of services selected at random ($\phi$). The comparison shows that the orders of convergence for the simulations are below those obtained with the analytic distribution $\bar{p}(q)$. This result agrees with the theoretical predictions for the distribution of X-motifs in the long term $p(q)$, which first-order dominates distribution $\bar{p}(q)$. The results also show the slow pace to zero of the mean number of X-motifs for low values of $\phi$, which will influence on the fit of the simulations to the theoretical distribution when the network size is not large enough.\par  
\vspace{0.5cm}
\noindent {\small Table 1: Order of convergence to zero of the mean number of X-motifs $<\bar{q}>$ predicted by the analytic results in equations (\ref{pqbar1}) and (\ref{pqbar2}) and of the mean number of X-motifs $<q>$ given by simulations of the model for a range of values of $\phi$. The parameters for the simulations are: $c=10$, $m=5$, $H_0=2$, $S_0=10$, the initial bipartite network is fully connected, stop time $t_F=5 \cdot 10^{3}$.}\par
\begin{center}
\begin{tabular}{ccc}

\hline
$\phi$ & analytic & simulations \\
 \hline
0  & 0 & 0  \\
0.1  & 0.11 & 0.07  \\
0.2  & 0.25 & 0.16  \\
0.3  & 0.43 & 0.36  \\
0.4  & 0.67 & 0.42  \\
0.5  & 1 & 0.57  \\
0.6  & 1 & 0.75  \\
0.7  & 1 & 0.78  \\
0.8  & 1 & 0.84  \\
0.9  & 1 & 0.79  \\
1  & 1 & 0.79  \\
\hline

\end{tabular}
\end{center}

\section{Simulations and comparison with real data}

\subsection{Study site and data}
In this section the model above is simulated and results are compared with the analytic estimations and real data. This data are extracted from the tourist activity developed in the southern area of the island of Gran Canaria, Spain. The zone entails six adjacent beaches along approximately 12 kms of coastal line, although is termed in the tourist market as the largest one, Maspalomas. \par

Tourism in Maspalomas started in the 60's, when the first hotels where built in a remote and rural area dedicated mainly to agriculture. The destination has maintained an irregular but marked growth in supply and visitors. Nowadays is a consolidated and mature destination hosting around three million tourists every year in approximately 700 accommodation units, composed mainly by hotels, apartments and bungalows.\par

The data to build the lodging-service network was collected from the user opinions published in the  web-site trypadvisor.com. This is a site for flight and accommodation reservations extended in most of the tourist destinations around the globe and the most popular ones among tourists\footnote{$https://www.tripadvisor.com/PressCenter$-$c4$-$Fact_Sheet.html$}. The registered users can publish opinions on lodgings and different services that are included in the web-site and assigned to a specific destination. Then, we consider that a tourist hosted in a lodging has visited a specific service if he/she has published and opinion of both elements in a time period of 15 days. The behavior of the theoretical representative tourist of a specific lodging is the result of adding all guests' behavior. By doing so, the scale effect of lodgings' capacity is included.\par

The data has been collected from September 2002 to March 2016. It includes 657 lodgings, 3065 services and 29,234 registered users that have made an opinion on a certain lodging and at least one service in Maspalomas in the prefixed time period. Lodgings include hotels, appartments and bungallows, given aside Bed\&Breakfast and other unofficial accommodation units. \par
%
\begin{figure}[!ht]
\centering
    \includegraphics[width=0.6\textwidth, height=0.4\textheight]{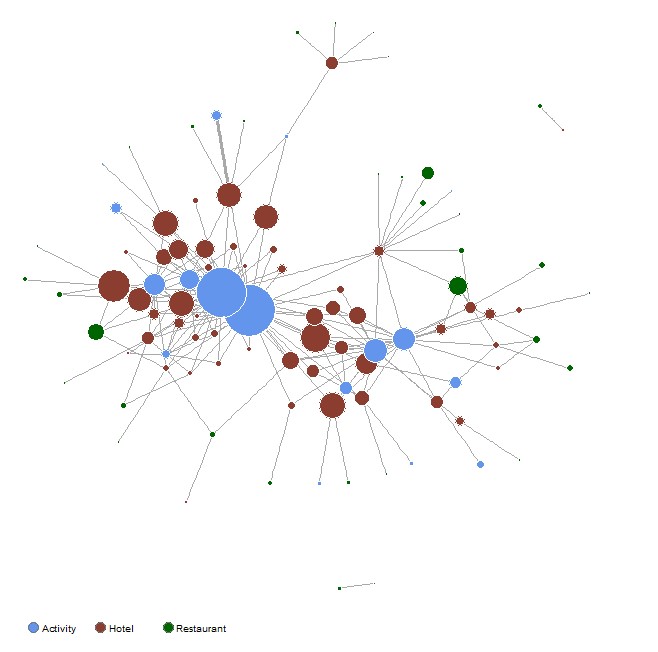}
\caption{Representation of the lodging-services network in Maspalomas (Gran Canaria). Lodgings are colored in brown and services in green (restaurants) and blue (rest of activities in the destination). The link lodging-service indicates that at least 15 opinions of the service was made by tourists hosted in the lodging. Node’s degree is represented by the ball size. Edges’ thickness illustrate the number of opinions. The spatial disposition of the networks was made using the Fruchterman-Reingold layout algorithm implemented in the R package igraph \cite{igraph}.}
\end{figure}
\vspace{0.2cm}

Figure 2 shows the lodging-service network in Maspalomas, where an heterogeneous degree distribution of both node categories can be observed. Isolated lodgings and services are excluded from the sample. So, we leave aside an amount of 73.3\% of lodgings and 51.2\% of services which did not receive any opinion. Table 2 includes the sample statistics.  The last parameter in Table 2 indicates that, in mean terms, around 30.8\% of the total links from a hotel to a service is part of an X-motif.\par
\vspace{0.5cm}
\noindent {\small Table 2: Basic statistics of the lodging-service network in Maspalomas. $H \equiv$ Logings; $S \equiv$ Services; $L \equiv$ Links; $<c> \equiv$ Mean degree of lodgings; $<s> \equiv$ Mean degree of services; $\rho \equiv$ Density; $<q/\sum_{i=1}^{s_j} (c_{j_i}-1)>$, where $c_{j_i}$ is the degree of lodging $i$ linked to service $j$.}
\begin{center}
\begin{tabular}{ccccccc}

\hline
H & S & L & $<c>$ & $<s>$ & $\rho$ & $<\frac{q}{\sum_{i=1}^{s_j} (c_{j_i-1)}>}$ \\
 \hline
182  & 1,496 & 13,359 & 73.4 & 8.9 & 9.5 $10^{-3}$ & 0.308\\
\hline

\end{tabular}
\end{center}
\vspace{0.2cm}

\subsection{Degree distributions}

In this section we compare the simulated, theoretical and empirical degree distributions. By hypothesis, the model assumes a constant degree of lodgings equal to $c$. However, Figure 3 shows that the empirical degree of lodgings is not constant, but follows a heterogeneous distribution. In order to compare the rest of distributions, we have assumed a lodging's degree in the model similar to the mean empirical degree, so $c=73$.\par

\begin{figure}[!ht]
\centering
    \includegraphics[width=0.8\textwidth, height=0.4\textheight]{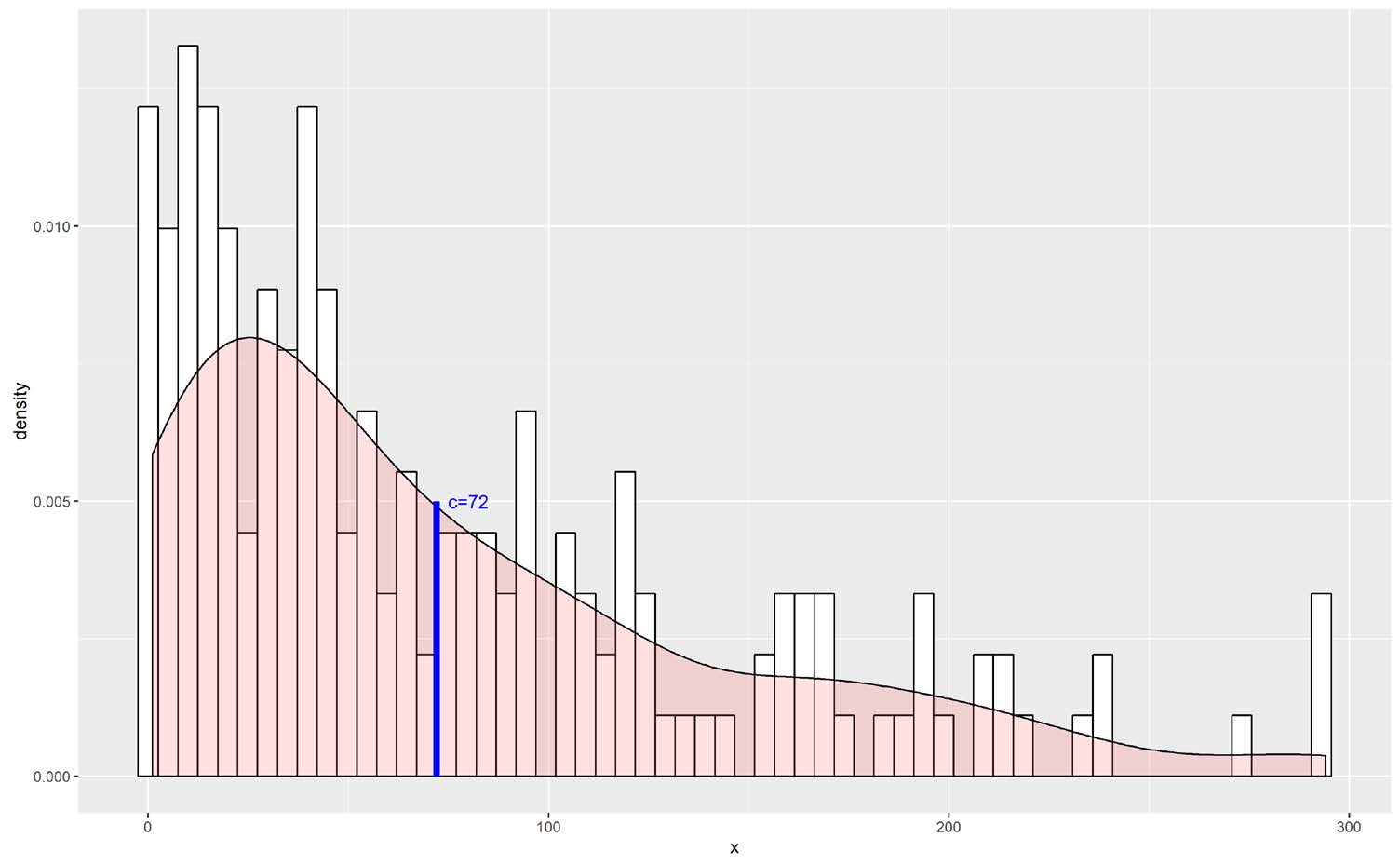}
\caption{Lodgings degree distribution in the empirical sample of Maspalomas (Gran Canaria). We have selected a constant degree $c=73$ in the simulations, which is close to the empirical mean degree.}
\end{figure}


Figure 4 shows the cumulative degree distribution of services. Every graph assumes a specific value of the percentage of services chosen at random in the model ($\phi$). As it can be observed, there is a good agreement of theoretical and simulated data in the middle range of $s$ values, before the cutoff in the simulated data.  The model predictions fits well the empirical results for a value $\phi=0.5$. The numerical simulations do not reach high values of $s$ for this time horizon due to the presence of services with zero links, which are excluded in the real sample. These results show that in this specific example the constant lodging's degree restriction does not limit the model capacity to represent the services' degree of the real data.\par  

\begin{figure}[!ht]
\centering
    \includegraphics[width=0.8\textwidth, height=0.4\textheight]{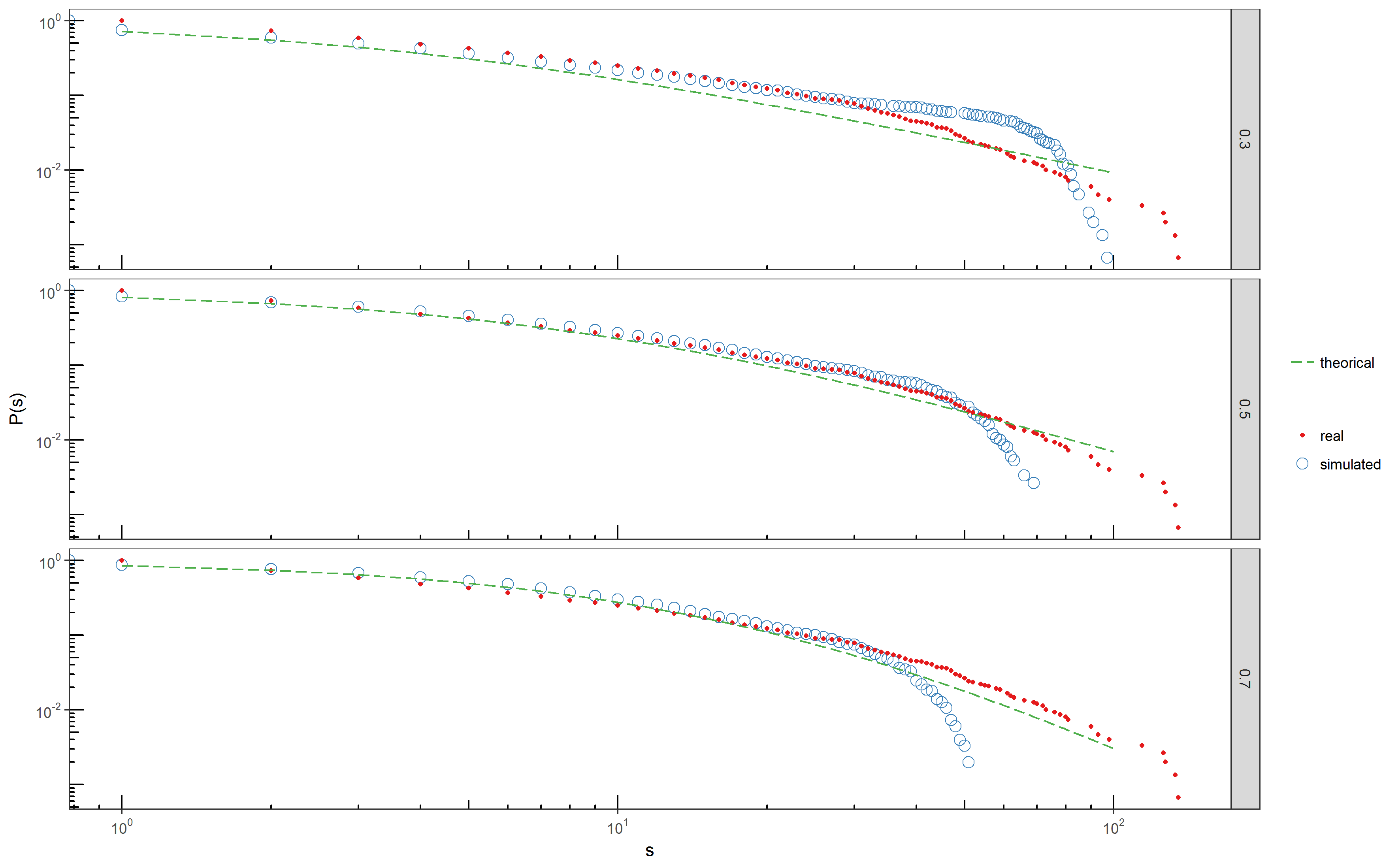}
\caption{Comparison among the cumulative degree distribution of services ($s$) for the empirical sample of recommendations in Maspalomas (red points), the simulations of the model (blue circles) and the theoretical predictions in Equation (\ref{dist_s}) (dashed green lines). Every graph assumes a different value of the percentage of services chosen at random: $\phi=0.3$ (up), $\phi=0.5$ (middle), $\phi=0.7$ (down). The other parameters are determined according the real data ($c$=73, $m$=8, $H_0$=2, $S_0$=56). The time horizon is $T=180$. A rather good agreement of the model with data is presented for $\phi=0.5$.}
\end{figure}


The statistics of the simulated network are close to the ones obtained with empirical sample (Table 2). By construction, $<c>$, $<s>$ and $\rho$ are very similar to the ones in Table 2. Moreover, the mean percentage of X-motif is $0.273$, which is proximate to the figure obtained with the real data.\par  

Figure 5 shows the cumulative degree distribution of services in the one-mode projection, assuming $\phi=0.5$. In this case, numerical simulations follow the shape of the empirical degree distribution, although do not match it. The theoretical predictions fail to represent the real data. In part, this may be a consequence of the high variance of the empirical degree distribution of lodgings (Figure 2), contrary to the model hypothesis of constant $c$. Nevertheless, other effects, not included in the model hypotheses, may also influence in the disagreement between analytic and empirical estimations. Moreover, it can be observed that the numerical degree distribution shows an uneven trajectory, which was not present in the two-mode degree distribution (Figure 4). 
This result illustrates the role of the low order of convergence to zero of the X-motifs on the agreement of the theoretical distribution to simulations and real data. 

\begin{figure}[!ht]
\centering
    \includegraphics[width=0.8\textwidth, height=0.4\textheight]{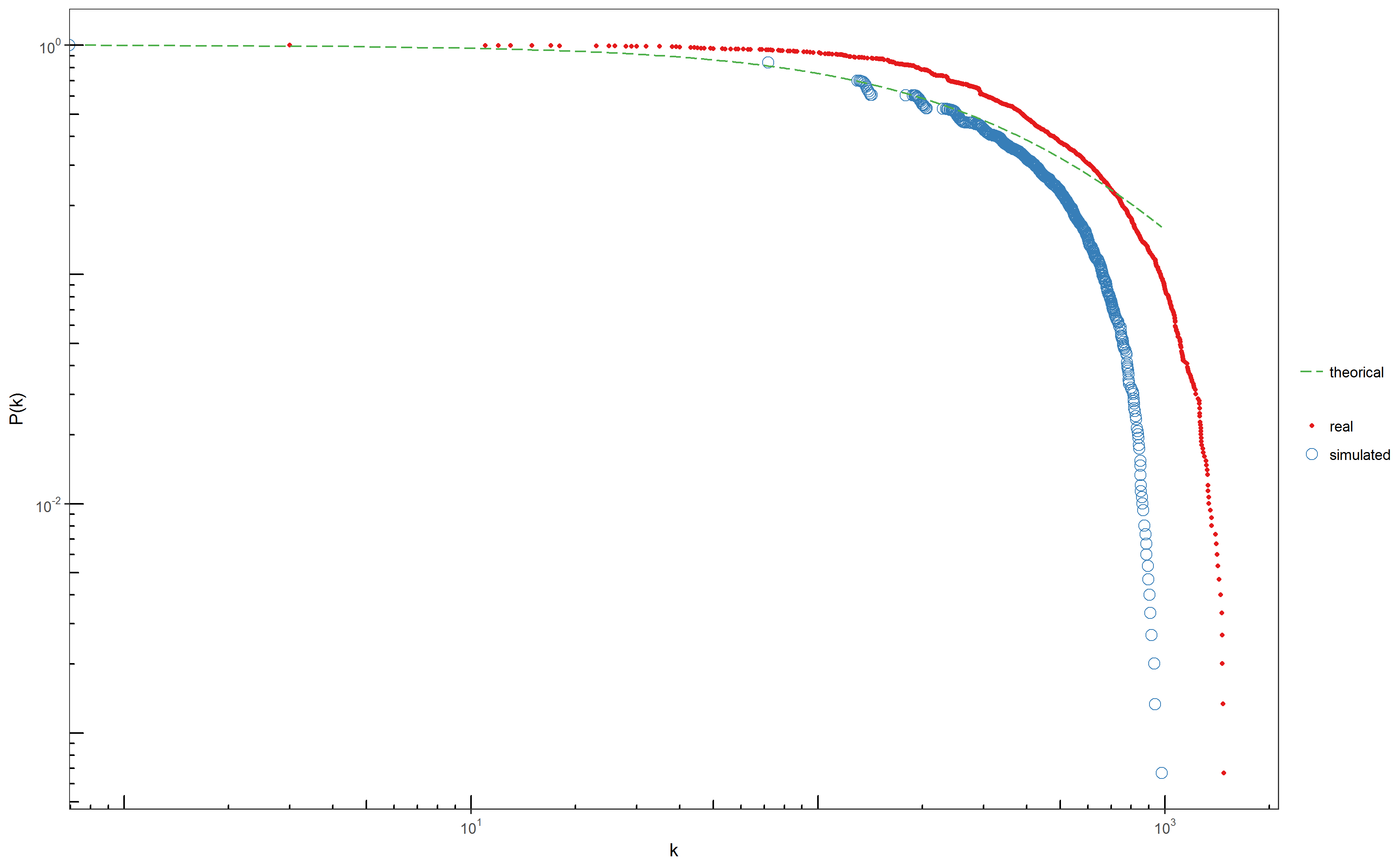}
\caption{Comparison among the cumulative degree distribution of services in the one-mode projection ($k$) for the empirical sample of recommendations in Maspalomas (red points), the simulations of the model (blue circles) and the theoretical predictions in Equation (\ref{dist_k}) (dashed green lines). The parameters are determined according the real data and the best agreement found in the two-mode network ($c$=73, $m$=8, $H_0$=2, $S_0$=56, $\phi=0.5$). The time horizon for the simulations is $T=180$.}
\end{figure}

\section{Conclusions}

In this paper, we have built an evolving bipartite network model that represents the supply network composed by the lodging and services in a tourism destination. These nodes are linked if a tourist hosted in a lodging enjoys a service during the stay. The model assumes that tourist's propensity to enjoy a service follow a combination of random and preferential attachment rule. These hypotheses are similar to other previous evolving bipartite network models. The model proposed here adapts some of the general conditions made in previous contributions to the specific case of tourists' behavior in a destination. The analytic results for the degree distributions of the bipartite network present good agreement of theoretical and simulation results. In the one-mode projection, some disturbances are produced between numerical and analytic results, due to the low order of convergence to zero of X-motifs.\par

The theoretical and numerical results were compared with a data sample extracted from the users' opinions on lodgings and services in Maspalomas, a tourist area in the island of Gran Canaria, Spain. The comparison shows that, in general terms, the  model fits well real data for a specific combination of the random and preferential attachment rules.\par  

This is a first theoretical and empirical analysis of the structure of the lodging-service network in a tourism destination. The model provides a realistic approximation of the real supply network, but it includes several limitations: (i) Distance is not currently considered in the attachment rules, although it is a factor that influence on the tourist's decision to visit a specific service; (ii) Rewiring of established links is not allowed, what indicates that, once visited, a service is permanently visited by next  tourists hosted in a specific lodging: (iii) Removing of lodgings/services due to aging or closure is not included either. Our future research will face these limitations with new extensions of the model. Additionally, in order to further check the model reliability, new comparisons with data from larger destinations in terms of visitors, lodgings and services, should be made.

\section{Acknowledgements}

The work was financed by project ECO2014-59067-P from the Ministry of Economy and Competitiveness of the Government of Spain.



\begin{thebibliography}{10}

\bibitem{Newman2001d}
M.~E.~J. Newman, ``{From the Cover: The structure of scientific collaboration
  networks},'' {\em Proceedings of the National Academy of Sciences}, vol.~98,
  pp.~404--409, jan 2001.

\bibitem{Goyal2006a}
S.~Goyal, M.~J. {Van Der Leij}, and J.~L. Moraga-Gonz{\'{a}}lez, ``{Economics:
  An emerging Small World},'' {\em Journal of Political Economy}, vol.~114,
  no.~2, pp.~403--412, 2006.

\bibitem{Watts1998}
D.~J. Watts and S.~H. Strogatz, ``{Collective dynamics of 'small-world'
  networks.},'' {\em nature}, vol.~393, no.~6684, pp.~440--442, 1998.

\bibitem{Serrano2003}
M.~A. Serrano and M.~Bogu{\~{n}}{\'{a}}, ``{Topology of the world trade
  web.},'' {\em Physical review. E, Statistical, nonlinear, and soft matter
  physics}, vol.~68, no.~1 Pt 2, p.~015101, 2003.

\bibitem{Garlaschelli2005}
D.~Garlaschelli and M.~I. Loffredo, ``{Structure and evolution of the world
  trade network},'' in {\em Physica A: Statistical Mechanics and its
  Applications}, vol.~355, pp.~138--144, 2005.

\bibitem{Fagiolo2009}
G.~Fagiolo, J.~Reyes, and S.~Schiavo, ``{World-trade web: Topological
  properties, dynamics, and evolution},'' {\em Physical Review E - Statistical,
  Nonlinear, and Soft Matter Physics}, vol.~79, no.~3, p.~036115, 2009.

\bibitem{Souma2003}
W.~Souma, Y.~Fujiwara, and H.~Aoyama, ``{Complex networks and economics},''
  {\em Physica A: Statistical Mechanics and its Applications}, vol.~324,
  no.~1-2, pp.~396--401, 2003.

\bibitem{Inaoka2004}
H.~Inaoka, H.~Takayasu, T.~Shimizu, T.~Ninomiya, and K.~Taniguchi,
  ``{Self-similarity of banking network},'' {\em Physica A: Statistical
  Mechanics and its Applications}, vol.~339, no.~3-4, pp.~621--634, 2004.

\bibitem{Mizuno2014}
T.~Mizuno, W.~Souma, and T.~Watanabe, ``{The structure and evolution of
  buyer-supplier networks},'' {\em PLoS ONE}, vol.~9, no.~7, 2014.

\bibitem{Scott2008}
N.~Scott, C.~Cooper, and R.~Baggio, ``{Destination Networks. Four Australian
  Cases},'' {\em Annals of Tourism Research}, vol.~35, no.~1, pp.~169--188,
  2008.

\bibitem{Romeiro2010}
P.~Romeiro and C.~Costa, ``{The potential of management networks in the
  innovation and competitiveness of rural tourism: a case study on the Valle
  del Jerte (Spain)},'' {\em Current Issues in Tourism}, vol.~13, no.~1,
  pp.~75--91, 2010.

\bibitem{Shih2006}
H.~Y. Shih, ``{Network characteristics of drive tourism destinations: An
  application of network analysis in tourism},'' {\em Tourism Management},
  vol.~27, no.~5, pp.~1029--1039, 2006.

\bibitem{Baggio2007}
R.~Baggio, ``{The web graph of a tourism system},'' {\em Physica A: Statistical
  Mechanics and its Applications}, vol.~379, no.~2, pp.~727--734, 2007.

\bibitem{Baggio2010}
R.~Baggio and C.~Cooper, ``{Knowledge transfer in a tourism destination: the
  effects of a network structure},'' {\em The Service Industries Journal},
  vol.~30, no.~10, pp.~1757--1771, 2010.

\bibitem{Miguens2008}
J.~I.~L. Migu{\'{e}}ns and J.~F.~F. Mendes, ``{Travel and tourism: Into a
  complex network},'' {\em Physica A: Statistical Mechanics and its
  Applications}, vol.~387, no.~12, pp.~2963--2971, 2008.

\bibitem{Grama2014}
C.~N. Grama and R.~Baggio, ``{A network analysis of Sibiu County, Romania},''
  {\em Annals of Tourism Research}, vol.~47, pp.~77--95, 2014.

\bibitem{Sainaghi2014}
R.~Sainaghi and R.~Baggio, ``{Structural social capital and hotel performance:
  Is there a link?},'' {\em International Journal of Hospitality Management},
  vol.~37, pp.~99--110, 2014.

\bibitem{Smallwood2012}
C.~B. Smallwood, L.~E. Beckley, and S.~A. Moore, ``{An analysis of visitor
  movement patterns using travel networks in a large marine park, north-western
  Australia},'' {\em Tourism Management}, vol.~33, no.~3, pp.~517--528, 2012.

\bibitem{Baggio2016}
R.~Baggio and R.~Sainaghi, ``{Mapping time series into networks as a tool to
  assess the complex dynamics of tourism systems},'' {\em Tourism Management},
  vol.~54, pp.~23--33, 2016.

\bibitem{Baggio2014}
R.~Baggio, ``{Complex tourism systems: a visibility graph approach},'' {\em
  Kybernetes}, vol.~43, no.~3-4, pp.~445--461, 2014.

\bibitem{DaFontouraCosta2009}
L.~{da Fontoura Costa} and R.~Baggio, ``{The web of connections between tourism
  companies: Structure and dynamics},'' {\em Physica A: Statistical Mechanics
  and its Applications}, vol.~388, no.~19, pp.~4286--4296, 2009.

\bibitem{Barabasi1999}
A.-L. Barab{\'{a}}si and R.~Albert, ``{Emergence of scaling in random
  networks},'' {\em Science}, vol.~286, no.~October, pp.~509--512, 1999.

\bibitem{Dorogovtsev2013}
S.~Dorogovtsev and J.~Mendes, {\em {Evolution of networks: From biological nets
  to the Internet and WWW}}.
\newblock Oxford University Press, 2013.

\bibitem{Ramasco2004}
J.~J. Ramasco, S.~N. Dorogovtsev, and R.~Pastor-Satorras, ``{Self-organization
  of collaboration networks},'' {\em Physical Review E - Statistical,
  Nonlinear, and Soft Matter Physics}, vol.~70, no.~3 2, pp.~1--10, 2004.

\bibitem{Noh2004}
J.~D. Noh, H.-C. Jeong, Y.-Y. Ahn, and H.~Jeong, ``{Growing network model for
  community with group structure},'' pp.~1--6, 2004.

\bibitem{Nacher2009}
J.~C. Nacher, T.~Ochiai, M.~Hayashida, and T.~Akutsu, ``{A mathematical model
  for generating bipartite graphs and its application to protein networks},''
  {\em Journal of Physics A: Mathematical and Theoretical}, vol.~42, no.~48,
  p.~485005, 2009.

\bibitem{Tian2012}
L.~Tian, Y.~He, H.~Liu, and R.~Du, ``{A general evolving model for growing
  bipartite networks},'' {\em Physics Letters, Section A: General, Atomic and
  Solid State Physics}, vol.~376, no.~23, pp.~1827--1832, 2012.

\bibitem{Zhang2013}
C.~X. Zhang, Z.~K. Zhang, and C.~Liu, ``{An evolving model of online bipartite
  networks},'' {\em Physica A: Statistical Mechanics and its Applications},
  vol.~392, no.~23, pp.~6100--6106, 2013.

\bibitem{Zhang2015b}
D.~Zhang, M.~Dai, L.~Li, and C.~Zhang, ``{Distribution characteristics of
  weighted bipartite evolving networks},'' {\em Physica A: Statistical
  Mechanics and its Applications}, vol.~428, pp.~340--350, 2015.

\bibitem{Qiao2016}
J.~Qiao, Y.~Y. Meng, H.~Chen, H.~Q. Huang, and G.~Y. Li, ``{Modeling one-mode
  projection of bipartite networks by tagging vertex information},'' {\em
  Physica A: Statistical Mechanics and its Applications}, vol.~457,
  pp.~270--279, 2016.

\bibitem{Lew2006}
A.~Lew and B.~McKercher, ``{Modeling tourist movements: A local destination
  analysis},'' {\em Annals of Tourism Research}, vol.~33, no.~2, pp.~403--423,
  2006.

\bibitem{McKercher2008}
B.~Mckercher and G.~Lau, ``{Movement Patterns of Tourists within a
  Destination},'' {\em Tourism Geographies}, vol.~10, no.~3, pp.~355--374,
  2008.

\bibitem{Feller1971}
W.~Feller, {\em {An Introduction to Probability Theory and Its Applications}},
  vol.~2.
\newblock John Wiley {\&} Sons, Inc., 2nd~ed., 1971.

\bibitem{Ramsay2006}
C.~M. Ramsay, ``{The Distribution of Sums of Certain I.I.D. Pareto Variates},''
  {\em Communications in Statistics - Theory and Methods}, vol.~35, no.~3,
  pp.~395--405, 2006.

\bibitem{igraph}
G.~Cs{\'{a}}rdi and T.~Nepusz, ``{The igraph software package for complex
  network research},'' {\em InterJournal Complex Systems}, vol.~1695, p.~1695,
  2006.

\end{thebibliography}

\end{document}